\documentclass[preprint,aps,prb,draft,amsmath,amssymb,endfloats]{revtex4}

\usepackage{bm}

\def\sro{SrRuO$_3$}
\def\gma{Ga$_{1-x}$Mn$_x$As}
\def\lsgo{LaSrGaO$_4$}
\def\sigmaxx{$\sigma_{\mathrm{xx}}$}
\def\sigmaxy{$\sigma_{\mathrm{xy}}$}
\def\xx{_{\mathrm{xx}}}
\def\xy{_{\mathrm{xy}}}
\def\f{_{\mathrm{F}}}
\def\k{_{\mathrm{K}}}
\def\h{_{\mathrm{H}}}
\def\thetaF{$\theta_{\mathrm{F}}$}
\def\thetaK{$\theta_{\mathrm{K}}$}
\def\thetaH{$\theta_{\mathrm{H}}$}

\begin{document}

\baselineskip = 24 pt


\title{Determination of the infrared complex magneto-conductivity tensor in itinerant ferromagnets from Faraday and Kerr measurements}



\author{M.-H. Kim,$^1$ G. Acbas,$^1$ M.-H. Yang,$^1$ I. Ohkubo,$^2$ H. Christen,$^3$
D. Mandrus,$^3$ M.A. Scarpulla,$^4$ O.D. Dubon,$^4$ Z. Schlesinger,$^5$ P. Khalifah,$^6$ and J. Cerne$^1$}

\affiliation{$^1$Department of Physics, University at Buffalo, SUNY,
Buffalo, NY 14260, USA}

\affiliation{$^2$Department of Applied Chemistry, Univ. of Tokyo, Tokyo, Japan}

\affiliation{$^3$Oak Ridge National Laboratory, Materials Science and Technology Division, Oak Ridge, TN 37831, USA}

\affiliation{$^4$Department of Materials Science and Engineering,
and Lawrence Berkeley National Laboratory, University of
California, Berkeley, CA 94720, USA}

\affiliation{$^5$Department of Physics, University of California, Santa Cruz, CA 95064, USA}

\affiliation{$^6$Department of Chemistry, University of Massachussetts, Amherst, MA 01003, USA}


\date{cond-mat/0701158, submitted to Phys. Rev. B, January 2007}

\begin{abstract}
We present measurement and analysis techniques that allow the complete complex magneto-conductivity tensor
to be determined from mid-infrared (11-1.6~$\mu$m; 100-800~meV) measurements of the complex Faraday (\thetaF) and Kerr (\thetaK) angles.
Since this approach involves measurement of the geometry (orientation axis and ellipticity of the polarization) of transmitted and reflected light, no absolute transmittance or reflectance measurements are required. Thick film transmission and reflection
equations are used to convert the complex \thetaF\ and \thetaK\ into the complex longitudinal conductivity
\sigmaxx\ and the complex transverse (Hall) conductivity \sigmaxy.
\thetaF\ and \thetaK\ are measured in a \gma\ and \sro\ films.  The resulting \sigmaxx\ is compared to the
values obtained from conventional transmittance and reflectance measurements, as well as the results from
Kramers-Kronig analysis of reflectance measurements on similar films.
\end{abstract}



\maketitle


\section{Introduction}

Conventional dc Hall effect measurements have been essential in revealing the unusual
character of novel electronic materials including high
temperature superconducting cuprates (HTSC),\cite{dchall} diluted
magnetic semiconductors (DMS),\cite{ohno} and ruthenate perovskite (RP)\cite{klein3, kats}
materials. In many of these materials, the Hall angle \thetaH\ and transverse (Hall)
conductivity \sigmaxy\ provide information critical to understanding their
electronic properties. The frequency dependence of \thetaH\ and \sigmaxy\ is very sensitive
to the electronic structure, and in many cases exposes new insights that are hidden from the longitudinal
conductivity \sigmaxx\ that is measured by
conventional spectroscopy.  The mid-infrared (MIR: 11-1.6~$\mu$m; 100-800~meV) energy range is particularly interesting
in many of these materials. For example, the band structure of RP\cite{fang} and III-V(Mn)
DMS\cite{sinova, hankiewicz}
leads to predictions of strong spectral features in \sigmaxy\ in the MIR. In the electron-doped cuprate,
Pr$_{2-x}$Ce$_x$CuO$_4$, evidence of a spin density wave gap has been observed in the MIR behavior
of \sigmaxy.\cite{zimmers}

Reflection and transmission magneto-polarimetry measurements allow
one to determine the complex Faraday \thetaF\ and Kerr \thetaK\
angles, respectively. \thetaF\ (\thetaK) describes the change in
polarization of transmitted (reflected) light produced by a sample
in a magnetic field. Although, \thetaF\ and \thetaK\ are useful,
they depend on the optical geometry of the sample, such as the
thickness of the film and index of refraction of the substrate.
Furthermore, calculations of \thetaF\ and \thetaK\ depend on which
assumptions are made for the optical formulas, e.g., thin film,
thick film, and bulk. Theoretical models typically calculate
response functions such as \sigmaxy\ and \sigmaxx. Since these
conductivities are related to the electronic behavior in a more
fundamental way, it is useful to convert \thetaF\ and \thetaK\
into these more elementary quantities.

In principle one can determine \sigmaxx\ using conventional, polarization-insensitive
spectroscopy techniques such as Kramers-Kronig analysis of
reflectance measurements or analysis of transmittance and
reflectance measurements. These measurements, however, do not access \sigmaxy,
which is critical to understanding many novel materials. Although
conventional spectroscopic techniques may be experimentally simpler than
the Faraday/Kerr measurements presented in this paper, polarization
insensitive approaches to measuring \sigmaxx\ rely on absolute transmittance
and reflectance measurements, which can limit the accuracy of these techniques. On the other hand, since
\thetaF\ and \thetaK\ measurements are absolute measurements that
do not require normalization to a reference sample, the accuracy
of this technique can be very high. {\it In this paper,
we obtain the entire complex magneto-conductivity tensor by measuring the
geometry (orientation axis and ellipticity of the polarization) of reflected and transmitted light;
no absolute intensity measurements are required.} A key advantage of determining both
\sigmaxx\ and \sigmaxy\ from the same set of \thetaF\ and \thetaK\
measurements is that the behavior of \sigmaxx, which may already
be fairly well known, can provide a consistency test for \sigmaxy,
which typically is not well known in the MIR. For experimental systems such
as ours that are designed for magneto-polarimetry measurements,
obtaining \sigmaxx\ through \thetaF\ and \thetaK\ measurements is
experimentally more straightforward and more accurate than using conventional
polarization insensitive spectroscopic techniques.

Of course, this approach has limitations. First, one must sensitively
measure polarization changes produced in transmitted and reflected light by a
magnetic field. Second, the samples must produce measurable
\thetaF\ and \thetaK\ signals, which is not always the case. For
example, in ordinary metals such as Au as well as unconventional
metals such as high temperature superconductors \thetaK\ is very
small and would be difficult to measure accurately. Fortunately,
magnetic metals such as \sro\ and \gma\ produce \thetaK\ signals
that can be measured readily.

In this paper, we present a MIR polarimetry technique that
determines the entire complex conductivity tensor using \thetaF\ 
and \thetaK\ measurements.  We shall first
introduce a sensitive magneto-optical measurement system, next we
develop formulas for \thetaF\ and \thetaK\ in terms of \sigmaxx\
and \sigmaxy, and finally we shall test these techniques on \sro\ and \gma\ films.

\section{Experimental System}

When illuminating a material in a magnetic field using linearly
polarized light, the transmitted and reflected beams can be
modified in two ways: the plane of polarization can be rotated
with respect to the incident linear polarization and the beam may
acquire ellipticity.  The polarization change of the transmitted
light is characterized by the complex Faraday angle $\theta\f$,
the optical analog of the Hall angle
$\theta\h=\sigma\xy/\sigma\xx$, where \sigmaxx\ is the
longitudinal conductivity and \sigmaxy\ is the transverse (Hall)
conductivity. $\theta\f$ relates the magnitudes and phases of the
transmitted electric fields that are perpendicular ($t\xy$) and
parallel ($t\xx$) to the incident linear electric polarization,
which is along the $x$-direction in this case: $\tan\theta\f\equiv
t\xy / t\xx$ where $t\xx$ and $t\xy$ are the diagonal and
off-diagonal components of the complex magneto-transmission
tensor.  The complex polar Kerr angle $\theta\k$ describes the change in
the polarization of reflected light for near-normal incidence, with $\tan\theta\k\equiv r\xy
/ r\xx$, where  $r\xx$ and $r\xy$ are the diagonal and
off-diagonal components of the complex magneto-reflection tensor.
For small changes in the incident polarization,
$\text{Re}\left(\theta\f\right)$
($\text{Re}\left(\theta\k\right)$) is related to the rotation and
$\text{Im}\left(\theta\f\right)$
($\text{Im}\left(\theta\k\right)$) is related to the ellipticity
of the transmitted (reflected) beam's polarization.

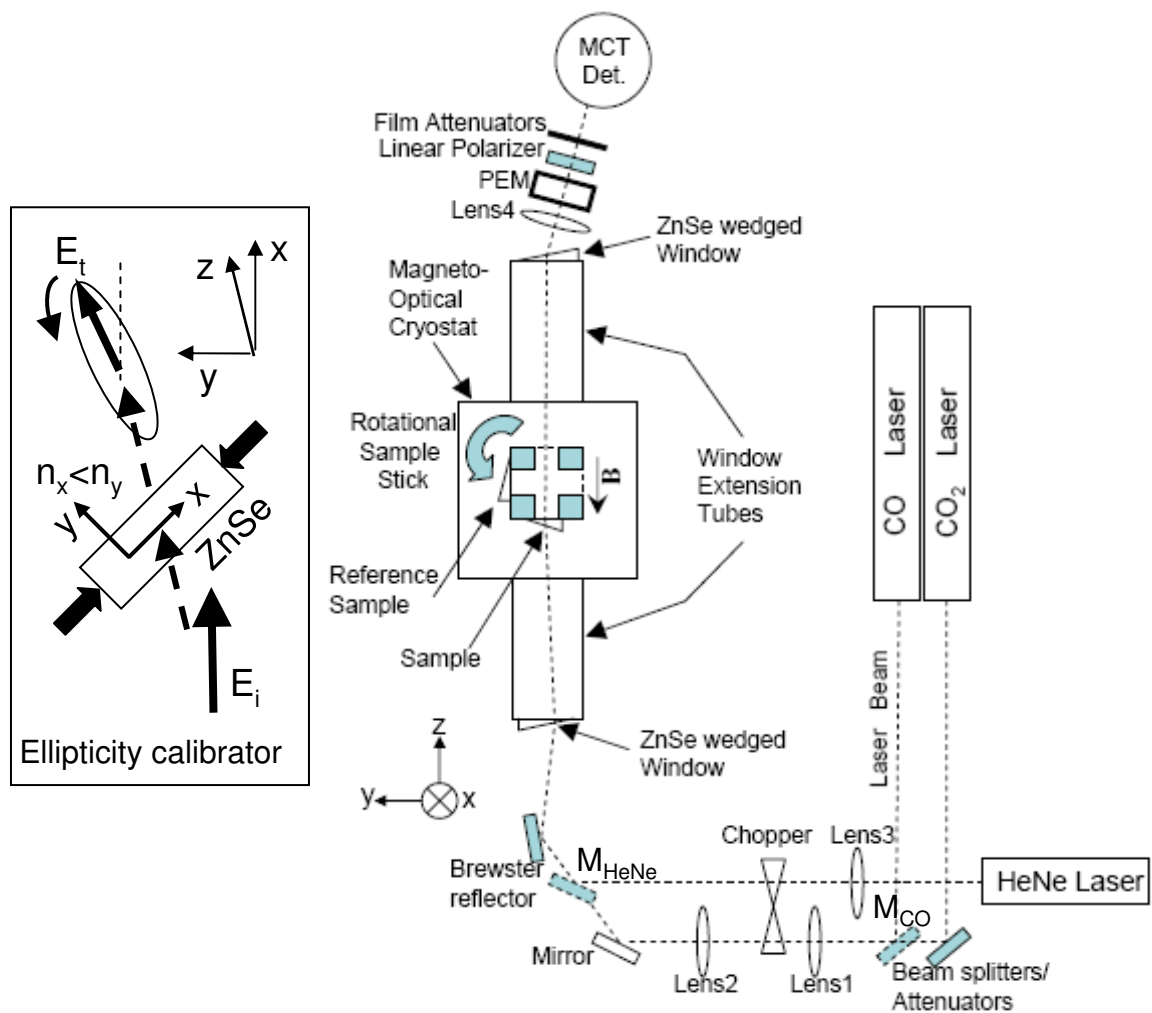
\begin{figure}
\caption{\label{fig;opticalPath} Overall schematic of the optical
path. The polarization change induced by compressively straining a ZnSe slide is shown in the boxed inset.}
\end{figure}

The experimental technique used in this paper is based on
Ref.~\onlinecite{cerne_rsi}. There are four new experimental
aspects that are added here: 1) precision translational mount for
a magneto-optical cryostat; 2) extended wavelength range using
additional lasers; 3) reflection measurements to determine
\thetaK; and 4) new calibration technique. After reviewing the
basic experimental technique, this paper will focus on these four
new areas. The experimental setup is shown in
Fig.~\ref{fig;opticalPath}. The Faraday and Kerr angles are
measured using discrete lines from CO$_{2}$ (115-133 meV), CO
(215-232meV), and HeNe (366 meV) lasers.  Measurements have also
been made using laser diodes operating at 500~meV and 775~meV
located near Lens 3 in Fig.~\ref{fig;opticalPath}. One can probe
the samples down to 6~K and magnetic fields up to 7 T in the
magneto-optical cryostat.  The different laser beams are aligned
in one optical path before entering the cryostat
(Fig.~\ref{fig;opticalPath}).  The original optical path is
designed for the CO$_{2}$ laser. The CO laser is sent to probe the
sample by inserting mirror $M_{\text{CO}}$, which is mounted on a
kinematic base before Lens 1.  The HeNe laser is sent to probe the
sample by inserting mirror $M_{\text{HeNe}}$, which is mounted on
a kinematic base near the Brewster reflector. An optical chopper
modulates the laser intensity with a frequency $\omega_{0}$. To
prevent etalon artifacts from multiple reflections within optical
components, the cryostat windows and the sample substrate are
wedged 1-2$^\circ$, and the photoelastic modulator (PEM), which is used to analyze
the polarization of the tranmsitted/reflected radiation, is
tilted forward 25$^\circ$. As is discussed in more detail in Section~\ref{sign_sect}, a
compressively strained ZnSe slide produces a well-characterized rotation and ellipticity
(see boxed inset in Fig.~\ref{fig;opticalPath}) in the polarization,
which can be used to determine the absolute signs of \thetaF\ and \thetaK.

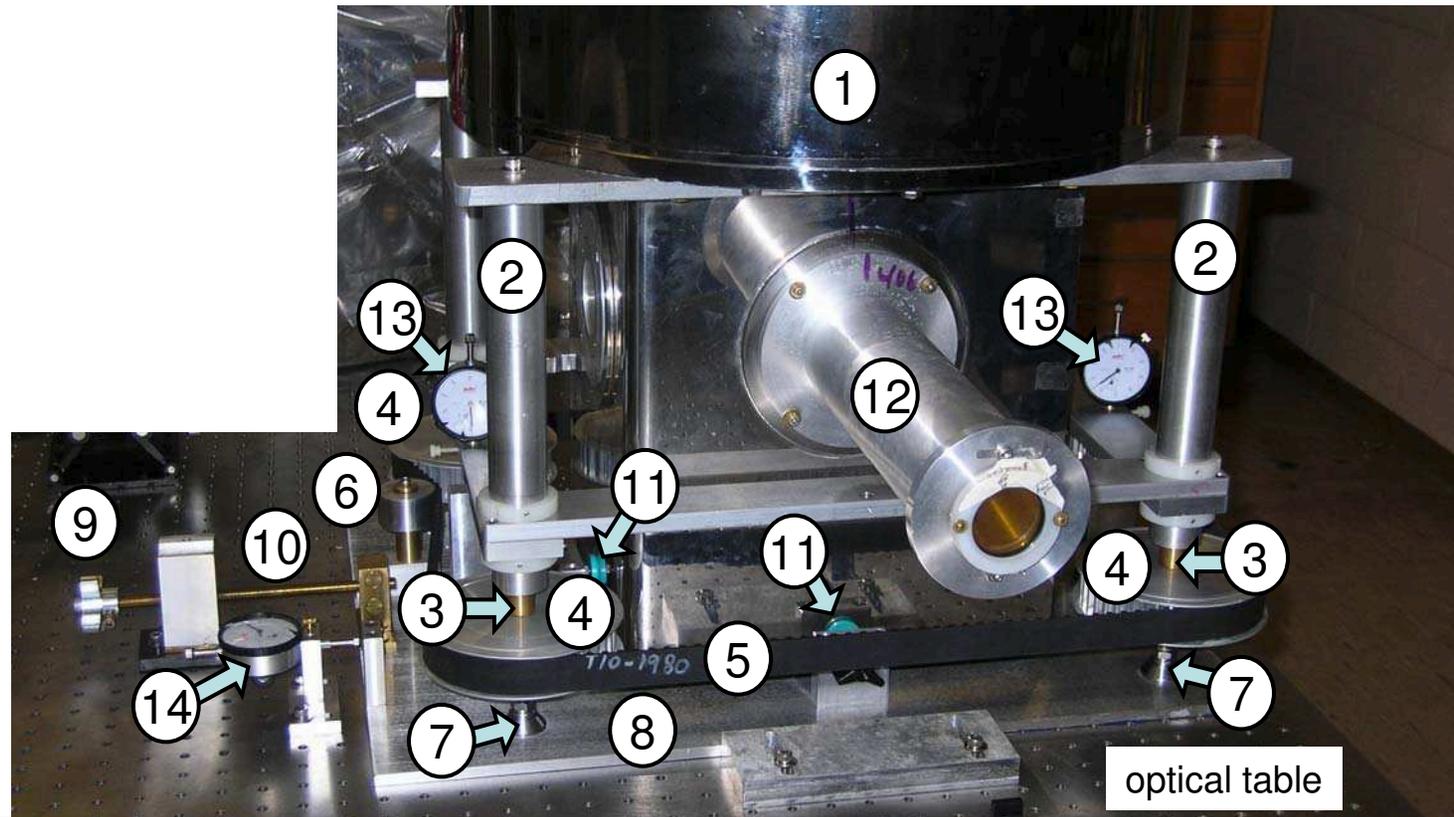
\begin{figure}
\caption{\label{fig;mount} The mount that allows the 700 pound magneto-optical cryostat (labeled 1 in figure)
to be translated vertically and horizontally with $\sim$~25~$\mu$m precision. The knob (9) attached to the
lead screw (10) for horizontal translation is indicated in the left part of the photograph.
Vertical adjustment is made by pulling on the timing belt (5), which rotates the four timing pulleys (4)
simultaneously, causing the large brass lead screws (3) to move the support legs (2) up or down.}
\end{figure}

The 7~T magneto-optical cryostat has only two windows, with the sample in vacuum. 
These room temperature ZnSe windows are placed on
30~cm extension tubes (as can be seen Fig.~\ref{fig;mount}) to
minimize their contribution to the Faraday rotation due to the
stray magnetic field. The absence of cold windows is important in
several respects. First, other magneto-optical cryostat can have
up to four cold windows (a pair of liquid helium temperature and a
pair of liquid nitrogen temperature windows) that are located
close to the sample and therefore experience similar magnetic
fields as the sample. Since the cold windows typically are several
millimeters thick, they can produce a large magneto-polarization
signal that can readily overwhelm the signal produced by the
sample, typically several hundred nanometer-thick films.
Furthermore, the absence of cold windows increases transmission
and reduces artifacts due to multiple reflection.

A photoelastic modulator (PEM) makes it possible to measure
sensitively both real and imaginary parts of \thetaF\ and
\thetaK.\cite{cerne_rsi} The optical axis of the PEM is oriented
vertically along the $x$ axis and modulates the phase of the
transmitted light that is polarized along the $y$-direction at a
frequency $\omega_{\text{PEM}}/2\pi\approx 50$~kHz. A linear
polarizer $P_2$ is oriented 45$^\circ$ from vertical and mixes the $x$
and $y$ polarization components of the light that has passed
through the PEM. A liquid-nitrogen-cooled
mercury-cadmium-telluride (MCT) detector measures the intensity of
the modulated beam.  If the sample does not cause any changes in
the incident polarization, the light entering the PEM is
linearly polarized along the $x$-direction and there will be no
signals at the detector related to the PEM. Three lock-in
amplifiers demodulate the detector signal. One lock-in amplifier
is referenced to chopper frequency $\omega_0$ to provide a measurement of the
average laser intensity $I_0$ at the detector. The other two
lock-in amplifiers are referenced to harmonics of $\omega_{\text{PEM}}$ to
detect the polarization of the beam. The even harmonics of
$\omega_{\text{PEM}}$ are related to a rotation of the
polarization vector [Re(\thetaF) or Re(\thetaK)] and the odd
harmonics are related to the ellipticity [Im(\thetaF) or
Im(\thetaK)].\cite{cerne_rsi} Typically, one measures the rotation
using the second harmonic signal $I_{2\omega_{\text{PEM}}}$ and the
ellipticity using the third harmonic signal
$I_{3\omega_{\text{PEM}}}$.

The mount holding the 7~T magneto-optical cryostat can be
translated with high accuracy ($\sim 25\ \mu$m) both horizontally and
vertically. As can be seen in Fig.~\ref{fig;mount}, the cryostat
rests on a large 5/16" thick aluminum base plate (17" by 24") that has
two long turcite (grade A, blue) strips attached to the bottom of
the base plate directly below the vertical supports. The strips are
0.030" thick, 23" long, and 2" wide, and are
epoxied into shallow pockets in the bottom of the plate.
Turcite is optimized for high load, low friction applications and does not
cold flow. The base plate and cryostat are moved horizontally with
respect to the optical table by turning a 3/8"x24 brass lead screw
that is attached to the base plate and the optical table. The height of
the cryostat is adjusted by simultaneously turning four 3/4"x16 brass lead screws
that are threaded into turcite bushings in each of the four
cryostat mount support legs. To insure that the cryostat is raised
and lowered uniformly, without tipping, the brass lead screws are
coupled to each other by a 1" wide, 80" long timing belt (T10,
10~mm pitch) that connects 5" diameter timing pulleys that are
attached to each lead screw. The design and construction of this
mount is challenging since it must be made of non-magnetic
materials and since the cryostat weighs approximately 700 pounds.
Separate leveling feet at the bottom of each brass lead screw
allow the cryostat to be leveled. The magneto-optical signals are
critically affected when the probe laser beam is close to the edge
of the aperture in the copper plate on which the sample is held, so centering the sample,
which can be as small as 3~mm, on the laser beam is very
important. Since the lasers are invisible, this can be difficult.
The procedure for centering the sample on the laser beam is
greatly improved and simplified using the translating cryostat
mount. The transmitted or reflected signal at the detector is
simply maximized by translating the sample (along with the
cryostat) vertically and horizontally. Precision indicators (shown in Fig.~\ref{fig;mount})
are placed to measure the absolute
position of the magneto-optical cryostat. By monitoring the
detector signal as one translates the cryostat and sample aperture
across the beam, one can determine the laser beam profile and
accurately center the sample on the beam. Since they
contain magnetic components, the indicators are removed before
the running measurements with the superconducting magnet energized. As with the
magneto-optical cryostat described in Ref.~\onlinecite{cerne_rsi},
unwanted motion of the sample is minimized by clamping the lower end of the sample tube to a horizontal delrin rod,
which is attached to an
aluminum plate in the side window opening of the tailpiece. Furthermore, the tailpiece of
the cryostat itself is clamped to the base plate by four bolts
that are capped with turcite pads, as shown in
Fig.~\ref{fig;mount}. This greatly increases the rigidity of the
cryostat mount and minimizes sample movement when the magnetic
field is energized.

Unlike Ref.~\onlinecite{cerne_rsi}, which describes measurements
in the 112-136~meV (1100-900~cm$^{-1}$; 11-9~$\mu$m) range, in
this paper the measurements have been extended up to 775~meV
(6250~cm$^{-1}$, 1.6~$\mu$m). Since the PEM, the lenses, and the
cryostat windows are ZnSe, the optical system is compatible with
sources over the 0.5-20~$\mu$m range. The main challenge in using
sources with shorter wavelength $\lambda$ is that the background
Faraday rotation due to stray magnetic fields at the windows and
sample substrate increases as $\lambda^{-2}$.\cite{ruymbeek} This
makes careful measurement and subtraction of the background at
shorter wavelengths especially critical. Since this background is
reduced when the energy gap of the optical material is increased
relative to the photon energy, higher band gap optical materials
such as BaF$_2$ and CaF$_2$ will be used for the cryostat windows
in future measurements. A further advantage of BaF$_2$ and CaF$_2$
is their smaller index (1.5 compared to 2.4 for ZnSe) that reduces
Fresnel losses in transmitting light through them. Initially, the
3.4~$\mu$m HeNe laser was placed approximately 1.5~m away from the magnet.
However the stray magnetic field at that distance induces both rotation
and ellipticity in the output of the HeNe laser that is comparable to
the signals produced by many of our samples (~$10^{-3}$~rad at 1~T).\cite{bedi}
This background is dramatically
reduced by moving the HeNe laser farther away (~2.5~m) from the
magneto-optical cryostat.

In this paper both transmission and reflection measurements are
made to determine both \thetaF\ and \thetaK, whereas only
transmission measurements (\thetaF) were reported in
Ref.~\onlinecite{cerne_rsi}. Although \thetaF\ and \thetaK\
provide similar information, there are several advantages to
measuring both. First, in the metallic films that are reported
here, the transmittance can be very low as is discussed in the
Section~\ref{samples} of this paper. The reflectance amplitudes, on the
other hand, are typically on the order of 50~\% or higher. Although the
substrates used in our measurements are relatively transparent in
the MIR, \thetaK\ measurements are not limited to transparent
substrates and can even be applied to bulk materials where
transmission measurements would be impossible for metals. The fact
that \thetaF\ and \thetaK\ are related is also an advantage in
that the self-consistency of the results may be confirmed.

In Ref.~\onlinecite{cerne_rsi} we describe several techniques to
calibrate the polarimetry system. We have developed a new
technique which allows the simultaneous calibration of the PEM
retardance $R_{\mathrm{PEM}}$ and the angle $\alpha_2$ of the
final linear polarizer $P_2$. One of the calibration techniques in
Ref.~\onlinecite{cerne_rsi} is to rotate the PEM by a known small
angle $\phi$ and use the change in the normalized detector signal
$S_2$ that is at $2\omega_{\mathrm{PEM}}$ to calibrate \thetaF.
The PEM and polarizer $P_2$ are rotated as a single unit, so that
the angle $\alpha_2$ between the PEM and $P_2$ is kept constant.
In this case, we measure the normalized signals $S_2$ and $S_4$ at
$2\omega_{\mathrm{PEM}}$ and $4\omega_{\mathrm{PEM}}$,
respectively. For $\phi\ll 1$ these signals depend on $\phi$ as
follows:

\begin{eqnarray}
S_2 &&= \frac{ I_{2\omega_{\mathrm{PEM}}} }   {I_0}
=\frac{ 4J_2( R_{\mathrm{PEM}} )\phi\tan(\alpha_2) } { 1+J_0(R_{\mathrm{PEM}})\phi\tan(\alpha_2) }\\
\label{eq;S2}
S_4 &&= \frac{ I_{4\omega_{\mathrm{PEM}}} }   {I_0}
=\frac{ 4J_4( R_{\mathrm{PEM}} )\phi\tan(\alpha_2) } { 1+J_0(R_{\mathrm{PEM}})\phi\tan(\alpha_2) }\\
\label{eq;S4}
\frac{S_2}{S_4}&&=\frac{ J_2( R_{\mathrm{PEM}} ) } {J_4( R_{\mathrm{PEM}} )}\hfill
\label{eq;Sratio}
\end{eqnarray}

\noindent where $J_n$ are $n$th order Bessel functions, $I_0$ is
the signal at the chopper frequency, $I_2\omega_{\mathrm{PEM}}$ is
the signal at $2\omega_{\mathrm{PEM}}$, and
$I_4\omega_{\mathrm{PEM}}$ is the signal at
$4\omega_{\mathrm{PEM}}$. Note that the ratio of $S_2$ and $S_4$
in Eq.~\ref{eq;Sratio} only depends on $R_{\mathrm{PEM}}$, as all
the other factors cancel. One can use the measurement of $S_2/S_4$
to determine $R_{\mathrm{PEM}}$. Since neither $\phi$ nor
$\alpha_2$  enter into Eq.~\ref{eq;Sratio}, this calibration is
not affected by the amount the PEM is rotated, as long $\phi\ll
1$, nor by the precise orientation of polarizer $P_2$. Once
$R_{\mathrm{PEM}}$ is determined, it can be entered into either
Eqs.~\ref{eq;S2}  or \ref{eq;S4} to determine $\alpha_2$, which is
nominally 45$^\circ$. Although $\alpha_2$ can be calibrated, when
doing measurements at a various wavelengths, errors can be reduced
by keeping $\alpha_2$ constant. The errors in adjusting $P_2$ to
make $\alpha_2=45^\circ$ at each wavelength can be significant. If
$\alpha_2$ is kept constant, there will be no variation from
wavelength to wavelength due to different settings of $P_2$. A
further check when performing this calibration at different
wavelengths and different lasers, is that $\alpha_2$ from the fits
should be the same since $P_2$ was not moved with respect to the PEM. In our calibration
measurements, $\alpha_2$ determined from this calibration
typically remains constant to within $0.2^\circ$ over the entire
measurement range. As in
Ref.~\onlinecite{cerne_rsi}, the roll-off attenuation of
the detector and its associated electronics is included in the
final calibration. We have found that the roll-off also depends on
the gain setting of the detector pre-amplifier.

\section{Samples}
\label{samples}

The \sro\ sample consists of a 282~nm thick \sro\ film on a \lsgo\
substrate and was grown by pulsed laser deposition at Oak Ridge
National Laboratory as described in Ref.~\onlinecite{khalifah2}.
The \lsgo\ substrate is transparent in the MIR, which allowed {\bf
both} transmission and reflection measurements to be made. The
thickness of the \sro\ film and the fact that the substrate begins
to absorb strongly below 117 meV, resulted in transmittances that
could be below 0.01~\%. Fortunately, high power coherent sources
such as the CO$_2$ and CO lasers coupled with a liquid nitrogen
cooled detector provide the measurement system with the necessary
dynamic range to measure small changes in the polarization even
for such small transmittance values. The back of the \lsgo\ substrate was polished
to a 2$^\circ$ wedge after the film was grown.

The \gma\ sample having a Curie temperature of 95~K was synthesized
using Mn$^+$ ion implantation followed by pulsed-laser
melting (II-PLM).\cite{scarpulla1,scarpulla2} A semi-insulating GaAs
(001) wafer was implanted with 80~keV
Mn$^+$ to a dose of $1.8\times10^{16}$~cm$^{-2}$ and irradiated in air
with a single 0.4~J/cm$^2$ pulse from a
KrF excimer laser.  The total Mn concentration depth profile measured
by secondary ion mass spectrometery (SIMS) was nearly Gaussian with a peak value of
approximately 8~\% and a width of 50~nm. The back of the GaAs substrate was polished
to a 1$^\circ$ wedge after the film was grown.

\section{Analysis}

Although many experimental improvements have been made since our
first report in
 Ref.~\onlinecite{cerne_rsi}, the primary contribution of this paper involves a novel analysis
of \thetaF\ and \thetaK\ measurements. In this section we develop
thick-film formulas for \thetaF\ and
\thetaK\ in terms of \sigmaxx\ and \sigmaxy. We also discuss the sign conventions that are used for \thetaF\ and \thetaK.

\subsection{Thick-film equations for \thetaF\ and \thetaK}

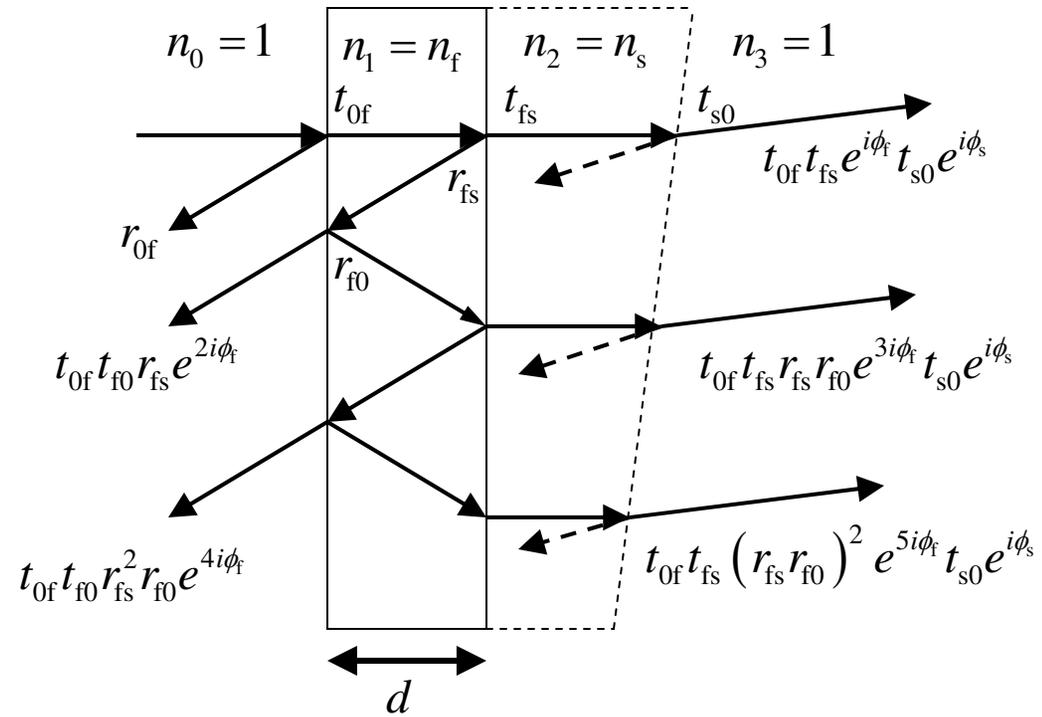
\begin{figure}
\caption{\label{fig;film} Optical path of light passing through a thick film on a wedged substrate. The multiply reflected beams in film are drawn at non-normal angles for clarity.}
\end{figure}

Since \thetaF\ and \thetaK\ are defined in terms of the complex
transmission and reflection amplitudes $t_\mathrm{xy}$,
$t_\mathrm{xx}$, $r_\mathrm{xy}$, and $r_\mathrm{xx}$, we begin by
determining these amplitudes. As light passes through a thick film
sample on a wedged substrate, part of the beam is reflected and
part is transmitted at each interface, as shown in
Fig.~\ref{fig;film}. The beam that reflects off the first air-film
interface combines with beams that have been multiply reflected
within the film to produce the reflected light. Since the back of
the substrate is wedged, beams that reflect from it do not combine
with the beams reflecting off the film. The first pass beam
combines with beams that are multiply reflected within the film to
produce the transmitted light entering the substrate. This
transmitted light is also multiply reflected within the substrate,
but since the substrate is wedged, each order of reflection exits
the substrate at a different angle. Therefore, the main advantage
of using a wedged substrate is that one can spatially separate the
first-pass beam from beams that are
 multiply reflected within the substrate.

Consider a thick film with complex index of refraction
$n_\mathrm{f}$ on a wedged substrate with complex index of
refraction $n_\mathrm{s}$. The Fresnel coefficients
$t_{\mathrm{ij}}$ and $r_{\mathrm{ij}}$ describe the transmission
and reflection amplitudes, respectively, at each interface
separating a material with index $n_\mathrm{i}$ from a material
with index $n_\mathrm{j}$. Summing over all transmitted beams
including only multiple reflections within the film, the complex
transmission coefficient for the sample is given by:

\begin{equation}
t (n_{\mathrm{f}}) = \frac{t_{\mathrm{0f}}t_{\mathrm{fs}} e^{i \phi_{\mathrm{f}}}}{1 - r_{\mathrm{fs}}
r_{\mathrm{f0}} e^{2 i \phi_{\mathrm{f}}}} t_{\mathrm{s0}}e^{i \phi_{\mathrm{s}}}
= \frac{2}{(1+n_{\mathrm{s}}) \cos{(kd)} - i
(n_{\mathrm{f}} + \frac{n_{\mathrm{s}}}{n_{\mathrm{f}}}) \sin{(kd)}} t_{\mathrm{s0}}e^{i \phi_{\mathrm{s}}},
\label{eq;t}
\end{equation}

\noindent where $k = \omega n_{\mathrm{f}} /c$ is the wave number
of the light within the film and $d$ is the thickness of the film.
The index 0 represents air. The phase shift of beam that passes
through the film is given by $\phi_{\mathrm{f}} = k d$.  For the
wedged substrate, we are only interested in the first pass beam,
which experiences a phase shift of $\phi_{\mathrm{s}} = (\omega
n_s d_{\text{sub}})/c$, where $d_{\text{sub}}$ is the thickness of
wedged substrate.

Similarly adding all contributions to the reflected light, the total reflection coefficient becomes:

\begin{equation}
r (n_{\mathrm{f}}) = \frac{r_{\mathrm{0f}} + r_{\mathrm{fs}} e^{2 i \phi_{\mathrm{f}}}}{1 - r_{\mathrm{fs}}
r_{\mathrm{f0}} e^{2 i \phi_{\mathrm{f}}}} = \frac{(n_{\mathrm{s}} - 1) \cos{(kd)} - i (n_{\mathrm{f}}
- \frac{n_{s}}{n_{f}}) \sin{(kd)}}{(1+n_{s}) \cos{(kd)} - i (n_{f}
+ \frac{n_{\mathrm{s}}}{n_{\mathrm{f}}}) \sin{(kd)}}.
\label{eq;r}
\end{equation}

\noindent Note that the $r (n_{\mathrm{f}})$ on a wedged substrate
is the same as for a film on an infinitely thick substrate since
the reflection from the back of the substrate never reaches the
detector.

The following calculation connects the \thetaF\ to the optical
conductivities \sigmaxx\ and \sigmaxy. We shall assign the
$z$-axis as the direction of light propagation and also the
direction of the magnetic field $B$. Assuming that the sample is
cylindrically symmetric along $z$-axis, the conductivity tensor
would be indistinguishable between $x$- and $y$-axis,
$\sigma_\mathrm{xx} = \sigma_\mathrm{yy}$ and $\sigma_\mathrm{xy}
= - \sigma_\mathrm{yx}$. Therefore, the optical conductivity
tensor is purely diagonal in the circular polarization basis.  One
can write the complex dielectric function $\varepsilon_\pm$ for a
circularly polarized basis with either positive or negative
helicity. $\varepsilon_\pm$ is related to the conductivity
$\sigma_\pm$, represented in the circular basis, by

\begin{equation}
\varepsilon_{\pm} = \varepsilon_\mathrm{b} - \frac{4 \pi}{i
\omega} \sigma_{\pm} = \varepsilon_\mathrm{b} - \frac{4 \pi}{i
\omega} (\sigma_\mathrm{xx} \pm i \sigma_\mathrm{xy}) =
\varepsilon_{0} \mp \frac{4 \pi}{\omega} \sigma_\mathrm{xy}
\label{eq;epsilon},
\end{equation}

\noindent where $\varepsilon_\mathrm{b}$ is the response of bound
charges and $\varepsilon_{0} \equiv \varepsilon_\mathrm{b} - 4 \pi
\sigma_\mathrm{xx} / i \omega $ is the longitudinal component of
dielectric function. The complex index of refraction
$n_{\mathrm{f},\pm}$ for the film using a circular polarization
basis can be expressed in terms of $\varepsilon_{\pm}$ and
$\sigma_\mathrm{xy}$.

\begin{equation}
n_{\mathrm{f},\pm} = \sqrt{\varepsilon_{\pm}} =
\sqrt{\varepsilon_{0}} \left( 1 \mp
\frac{\sigma_\mathrm{xy}}{\sigma_\mathrm{b} + i
\sigma_\mathrm{xx}} \right)^{1/2}, \label{eq;n}
\end{equation}

\noindent where $\sigma_\mathrm{b}$ is defined as $\frac{\omega}{4
\pi} \varepsilon_\mathrm{b}$.  Since $\sigma_\mathrm{xx} \gg
\sigma_\mathrm{xy}$ in most cases in the MIR, $n_{\mathrm{f},\pm}$
can be simplified to

\begin{equation}
n_{\mathrm{f},\pm} \simeq \sqrt{\varepsilon_0} \mp \frac{1}{2}
\sqrt{\varepsilon_0} \frac{\sigma_\mathrm{xy}}{\sigma_{\mathrm{b}}
+ i \sigma_\mathrm{xx}} = n_{\mathrm{f},0} \mp \delta
n_\mathrm{f}, \label{eq;nf}
\end{equation}

\noindent where $n_{\mathrm{f},0} = \sqrt{\varepsilon_{0}}$ is
longitudinal component of complex index of refraction and $\delta
n_\mathrm{f} = 2 \pi \sigma_\mathrm{xy} / \omega n_{\mathrm{f},0}$
is the transverse component.  The diagonal (longitudinal)
transmission coefficient $t_\mathrm{xx}$ and the off-diagonal
(transverse) transmission coefficient $t_\mathrm{xy}$ in a linear
polarization basis are related to the diagonal transmission
coefficients $t_+$ and $t_-$ in the circular polarization basis as
follows.  Since $t_\pm (n_{\mathrm{f},\pm}) \approx t_0
(n_{\mathrm{f},0}) \pm ({\partial t_0}/{\partial
n_{\mathrm{f},0}}) \delta n_{\mathrm{f}}$,

\begin{eqnarray}
  t_\mathrm{xx} &=& \frac{t_{+} + t_{-}}{2} \approx t_0 (n_{\mathrm{f},0}) , \label{eq;txx}\\
  t_\mathrm{xy} &=& \frac{t_{+} - t_{-}}{2 i} \approx \frac{1}{i} \left( \frac{\partial t_0}{\partial n_{\mathrm{f},0}}
\right) \delta n_\mathrm{f}. \label{eq;txy}
\end{eqnarray}

\indent Equations (\ref{eq;txx}) and (\ref{eq;txy}) show that
diagonal transmission measurements probe the sum of $t_+$ and
$t_-$, whereas off-diagonal transmission measurements, e.g.,
Faraday measurements, probe the difference, making them more
sensitive to small changes in $t_+$ and $t_-$ induced by magnet
fields or other symmetry-breaking mechanisms. Putting
Eqs.~(\ref{eq;txx}) and (\ref{eq;txy}) together, the Faraday angle
$\theta_{\text{F}}$ can be represented in terms of
$\sigma_\mathrm{xy}$ and $\sigma_\mathrm{xx}$ (found in
$n_{\mathrm{f},0}$) as

\begin{equation}
\tan \theta_\mathrm{F} = \frac{t_\mathrm{xy}}{t_\mathrm{xx}} = - i
\delta n_\mathrm{f} \frac{1}{t_0(n_{\mathrm{f},0})} \frac{\partial
t_0}{\partial n_{\mathrm{f},0}} = \frac{- 2 \pi i}{\omega
n_{\mathrm{f},0}} \left( \frac{1}{t_0(n_{\mathrm{f},0})}
\frac{\partial t_0}{\partial n_{\mathrm{f},0}} \right)
\sigma_\mathrm{xy}. \label{eq;tanF}
\end{equation}

\noindent One can use the complex transmission coefficient
$t_0$($n_{\mathrm{f},0}$) (Eq.~(\ref{eq;t})) to calculate the
Faraday angle $\theta_{\text{F}}$.
\begin{equation}\label{eq;Fara_thickfilm}
\tan \theta_\mathrm{F} = \left( \frac{-2 \pi i
\sigma_\mathrm{xy}}{\omega n_{\mathrm{f},0}} \right) \frac{ \left[
(1+n_s) \left( \frac{\omega d}{c} \right) + i \left( 1 -
\frac{n_s}{n_{\mathrm{f},0}^2} \right) \right] \sin (kd) + i
\left( n_{\mathrm{f},0} + \frac{n_s}{n_{\mathrm{f},0}} \right)
\left( \frac{\omega d}{c} \right) \cos (kd)}{(1+n_s) \cos (kd) - i
\left(n_{\mathrm{f},0} + \frac{n_s}{n_{\mathrm{f},0}}\right) \sin
(kd)}.
\end{equation}

\noindent Taking the approximation $kd = (2\pi d)/\lambda \ll 1$
($d \rightarrow$ 0, $\omega \rightarrow$ 0), the
Eq.~(\ref{eq;Fara_thickfilm}) yields the simple thin-film formula
\begin{equation}
\tan \theta_\mathrm{F} \approx \left(
\frac{\sigma_\mathrm{xy}}{\sigma_\mathrm{xx}} \right) \left[ 1 +
\frac{1}{Z_+ \sigma_\mathrm{xx}} \right]^{-1},
\label{eq;faraday_cond}
\end{equation}
where $Z_\pm \equiv (Z_0 d)/(n_s \pm 1)$, $Z_0$ is the impedance
of free space and the unit of conductivity is $\Omega^{-1}$
cm$^{-1}$. Note that Eq.~(\ref{eq;faraday_cond}) is slightly
different from Eq.~(2.2) in Ref.~\onlinecite{au} due to a typographical
error on the right side of Eq.~(2.2) .

One can use the same approach to calculate \thetaK\ in terms of
\sigmaxx\ and \sigmaxy\ using the reflection coefficients.  The
diagonal and the off-diagonal reflection amplitudes,
$r_\mathrm{xx}$ and $r_\mathrm{xy}$, can be expressed in a linear
polarization basis in terms of the diagonal reflection
coefficients $r_+$ and $r_-$ in the circular polarization basis.

\begin{eqnarray}
  r_\mathrm{xx} &=& \frac{r_{+} + r_{-}}{2} \approx r_0(n_{\mathrm{f},0}),  \label{eq;rxx}\\
  r_\mathrm{xy} &=& \frac{r_{+} - r_{-}}{2 i} = \frac{1}{i} \left( \frac{\partial r_0}{\partial n_{\mathrm{f},0}}
\right) \delta n_\mathrm{f} . \label{eq;rxy}
\end{eqnarray}

\noindent Again, we have assumed that $r_\pm (n_{\mathrm{f},\pm})
\approx r_0 (n_{\mathrm{f},0}) \pm ({\partial r_0}/{\partial
n_{\mathrm{f},0}}) \delta n_{\mathrm{f}}$.  Combining
Eq.~(\ref{eq;rxx}) and Eq.~(\ref{eq;rxy}) produces an expression
for \thetaK\ in terms of \sigmaxx\ and \sigmaxy,

\begin{equation}
\tan \theta_\mathrm{K} = \frac{r_\mathrm{xy}}{r_\mathrm{xx}} = - i
\delta n_\mathrm{f} \frac{1}{r_0(n_{\mathrm{f},0})} \frac{\partial
r_0}{\partial n_{\mathrm{f},0}} = \frac{- 2 \pi i}{\omega
n_{\mathrm{f},0}} \left( \frac{1}{r_0 (n_{\mathrm{f},0})}
\frac{\partial r_0}{\partial n_{\mathrm{f},0}} \right)
\sigma_\mathrm{xy}. \label{eq;tanK}
\end{equation}

\noindent
To calculate the Kerr angle $\theta_{\text{K}}$, we can
use the complex reflection coefficient $r_0$ (Eq.~(\ref{eq;r})).
\begin{eqnarray}\nonumber \label{eq;Kerr_thickfilm}
\tan &\theta_\mathrm{K}& = \left( \frac{-2\pi i
\sigma_\mathrm{xy}}{\omega n_{\mathrm{f},0}} \right) \frac{ \left[
-(n_s-1) \left( \frac{\omega d}{c} \right) - i \left( 1 +
\frac{n_s}{n_{\mathrm{f},0}^2} \right) \right] \sin (kd) - i
\left( n_{\mathrm{f},0} - \frac{n_s}{n_{\mathrm{f},0}} \right)
\left( \frac{\omega d}{c} \right) \cos (kd)}{(n_s - 1) \cos (kd) -
i \left( n_{\mathrm{f},0} - \frac{n_s}{n_{\mathrm{f},0}}\right)
\sin (kd)}
\\ &+& \left( \frac{-2\pi i \sigma_\mathrm{xy}}{\omega
n_{\mathrm{f},0}} \right) \frac{ \left[ (n_s+1) \left(
\frac{\omega d}{c} \right) + i \left( 1 -
\frac{n_s}{n_{\mathrm{f},0}^2} \right) \right] \sin (kd) + i
\left( n_{\mathrm{f},0} + \frac{n_s}{n_{\mathrm{f},0}} \right)
\left( \frac{\omega d}{c} \right) \cos (kd)}{(n_s + 1) \cos (kd) -
i \left( n_{\mathrm{f},0} + \frac{n_s}{n_{\mathrm{f},0}}\right)
\sin (kd)}
\end{eqnarray}

\noindent Applying the same approximation ($kd \ll 1$) used for
\thetaF\ to Eq.~(\ref{eq;Kerr_thickfilm}) results in the thin-film formula for
\thetaK,
\begin{equation}
\tan \theta_\mathrm{K} \approx \left(
\frac{\sigma_\mathrm{xy}}{\sigma_\mathrm{xx}^2} \right) \left(  -
\frac{2}{Z_0 d} \right) \left[ \left( 1+ \frac{1}{Z_+
\sigma_\mathrm{xx}} \right) \left( 1+ \frac{1}{Z_-
\sigma_\mathrm{xx}} \right) \right]^{-1}, \label{eq;kerr_cond}
\end{equation}
where the unit of conductivity is $\Omega^{-1}$ cm$^{-1}$.

Equations (\ref{eq;tanF}) and (\ref{eq;tanK}) can be simplified by
using the relation between $\sigma_\mathrm{xx}$ and
$n_{\mathrm{f},0}$ obtained by the Eqs.~(\ref{eq;epsilon}) and
(\ref{eq;nf}).
\begin{eqnarray}\nonumber
\tan \theta_\mathrm{F} &=& -\frac{1}{t_0} \left( \frac{\partial
t_0}{\partial \sigma_\mathrm{xx}} \right) \sigma_\mathrm{xy}, \\
\tan \theta_\mathrm{K} &=& -\frac{1}{r_0} \left( \frac{\partial
r_0}{\partial \sigma_\mathrm{xx}} \right) \sigma_\mathrm{xy}.
\end{eqnarray}

Note that since both $\tan \theta_\mathrm{F}$ and $\tan
\theta_\mathrm{K}$ are proportional to $\sigma_\mathrm{xy}$, the
magneto-optical signals vanish when $\sigma_\mathrm{xy} = 0$, as
expected.  Dividing $\tan\theta\f$ in Eq.~(\ref{eq;tanF})
$\tan\theta\k$ in Eq.~(\ref{eq;tanK}) allows $\sigma_\mathrm{xy}$
to divide out and produces

\begin{equation}
\frac{\tan \theta_\mathrm{F}}{\tan \theta_\mathrm{K}} =
\frac{\frac{1}{t_0} \left( \frac{\partial t_0}{\partial
n_{\mathrm{f},0}} \right)} { \frac{1}{r_0} \left( \frac{\partial
r_0}{\partial n_{\mathrm{f},0}} \right)} = \text{F}
(n_{\mathrm{f},0}). \label{eq;ratioFaraKerr}
\end{equation}

\noindent
Also, it can be expressed as
\begin{equation}
\frac{\tan \theta_\mathrm{F}}{\tan \theta_\mathrm{K}} =
\frac{\frac{1}{t_0} \left( \frac{\partial t_0}{\partial
\sigma_\mathrm{xx}} \right)} { \frac{1}{r_0} \left( \frac{\partial
r_0}{\partial \sigma_\mathrm{xx}} \right)} = \text{G}
(\sigma_\mathrm{xx}).
\end{equation}

\noindent Since $\delta n_\mathrm{f} \ll n_{\mathrm{f},0}$, $t_0
\approx t_\mathrm{xx}$ and $r_0 \approx r_\mathrm{xx}$ in the linear
polarization basis, and the complex function $\text{F}
(n_{\mathrm{f},0})$ only depends on the longitudinal index of
refraction of the film. If the complex \thetaF\ and \thetaK\ are
measured experimentally, one can solve
Eq.~(\ref{eq;ratioFaraKerr}) numerically to obtain
$n_{\mathrm{f},0}$. Once $n_{\mathrm{f},0}$ is determined, we can
use Eqs.~(\ref{eq;epsilon}) and (\ref{eq;nf}) to calculate the
complex longitudinal conductivity $\sigma_\mathrm{xx}$ of the
film:

\begin{equation}
\sigma_\mathrm{xx} = \frac{i \omega}{4 \pi}
(\varepsilon_{\mathrm{b}} - n_{\mathrm{f},0}^2),
\end{equation}

\noindent where $\varepsilon_\mathrm{b}$ is the contribution to
the dielectric function from bound carriers, which allows the
conductivity to be determined for free carriers. In these
measurements, we typically are interested in the response of all
the carriers, bound and free, so $\varepsilon_\mathrm{b}$ is set
to 0. Using the measured values for \thetaF\ or \thetaK, and
plugging $n_{\mathrm{f},0}$ back into Eqs.~(\ref{eq;tanF}) or
(\ref{eq;tanK}), one can now determine the complex
$\sigma_\mathrm{xy}$.

\subsection{Sign calibration}
\label{sign_sect}

The sign calibration is important for our Faraday and Kerr measurements
because the solutions for \sigmaxx\ and \sigmaxy\ depend on which
signs are assigned to \thetaF\ and \thetaK.  There are two sign
conventions in describing the time-evolution of the electric field
in magneto-optical measurements: $\exp(-i\omega t)$ or
$\exp(+i\omega t)$. Determining the correct signs for \thetaF\ and
\thetaK\ can be challenging, both experimentally and
theoretically.

Experimentally, one must make sure that the direction of the
magnetic field is known and that the direction of the changes in
orientation and  ellipticity in the polarization of the
transmitted and reflected light are properly determined. Since the
signals are demodulated using lock-in amplifiers, one must keep
track of the phase of each lock-in in order to avoid sign errors.

In our measurements the signs of the polarization signals are
determined in three independent, yet overlapping, ways. The sign
of the polarization rotation ($\text{Re}\left(\theta\f\right)$ and
$\text{Re}\left(\theta\k\right)$) is determined by rotating the
PEM and linear polarizer $P_2$ together in a counter-clockwise
direction as viewed along the beam's propagation direction towards
the detector. This is equivalent to the sample rotating the
transmitted/reflected polarization in the clockwise direction. The
change in signal is compared with that produced by the sample in a
magnetic field. Second, the sign of both the rotation and
ellipticity can be verified by placing a ZnSe slide in the beam as
described in Refs.~\onlinecite{schmadel_thesis,cerne_rsi}. The
index of refraction in ZnSe is decreased for the linear
polarization along the compressive strain direction,\cite{yu} and
hence this polarization exits the ZnSe slide before the
polarization that is perpendicular to the compressively strained
axis. For the geometry shown in the boxed inset of
Fig.~\ref{fig;opticalPath}, this phase shift between perpendicular
linear polarizations produces a counterclockwise rotation of the
linear polarization as well as a counterclockwise ellipticity, as viewed towards the detector. The
strain on the ZnSe slide is applied by hanging a weight on a
compression lever, as can be seen in greater detail in
Ref.~\onlinecite{schmadel_thesis}. Since the orientation of the slide
critically affects the tranmsitted polarization, the slide holder
is clamped to the optical table to prevent small movements of the
slide when the weight is added or removed. The same calibration
can be made using a waveplate, as described in
Ref.~\onlinecite{cerne_rsi}, but a ZnSe slide is much less
expensive than a zeroth order infrared waveplate, and since the
strain is applied externally, there is no ambiguity for the
direction of the fast axis, which is not always clear on a
waveplate. The compressed slide sign calibration technique can be
used over a large wavelength range (500~nm to 20~$\mu$m) without
any sign changes for two reasons: 1) the strain, and therefore the
retardance, of the slide is small, so even at the shortest
wavelengths in this range the retardance never apporoaches $\pi$,
where fast and slow axes would reverse. 2) the piezobirefringence
coefficient for ZnSe does not change sign in this wavelength
range.\cite{yu} These sign calibrations are performed with the
sample in place and under exactly the same conditions that are used
for measuring the sample. Once the calibrations are completed, the
PEM is aligned with the laser polarization and the ZnSe slide is
removed before the magnetic field is energized. Once the
directions of the rotation and ellipticity signals are determined
with respect to the magnetic field direction and the direction of
light propagation, one can use Ref.~\onlinecite{sign} to determine
the proper signs of \thetaF \ and \thetaK.

Finally, the signs of polarization signals are verified by 
measuring $\text{Re}\left(\theta\f\right)$ and $\text{Im}\left(\theta\f\right)$ produced by a gold film at
120~meV. Since \thetaF\ in gold is produced by free electrons, one
can characterize the response of a sample in
$\text{Re}\left(\theta\f\right)$ and
$\text{Im}\left(\theta\f\right)$ as ``electron-like"
or ``hole-like."  One can represent the complex \sigmaxx\ and
\sigmaxy\ of gold using a simple Drude model, which in turn can be
used to calculate \thetaF\ using the thick-film transmission
formula in Eq.~(\ref{eq;tanF}). This same formula is then used to calculate ``backwards" from the measured \thetaF\ and \thetaK\ produced by other samples to obtain \sigmaxx\ and \sigmaxy. The signs of
$\text{Re}\left(\theta\f\right)$  and
$\text{Im}\left(\theta\f\right)$ determined by this calculation
can be used to determine the signs produced by other samples. For
example, applying Eq.~(\ref{eq;tanF}) to a Drude model for a gold
film produces $\text{Re}\left(\theta\f\right) <0$ and
$\text{Im}\left(\theta\f\right)<0$ below 200~meV. Therefore, if the
$\text{Re}\left(\theta\f\right)$ signal for a sample has the
same polarity with applied magnetic field as the signal from a
gold film (electron-like), a negative value of
$\text{Re}\left(\theta\f\right)$  is used in the thick-film
equation to determine \sigmaxx\ and \sigmaxy. The sign of
$\text{Im}\left(\theta\f\right)$ is determined the same way. This
convention insures that a film with an electron-like response in
both $\text{Re}\left(\theta\f\right)$  and
$\text{Im}\left(\theta\f\right)$ will produce the correct signs
and magnitudes of \sigmaxx\ and \sigmaxy. Since \thetaK\ from a gold
film is too small to measure, we cannot us the gold film to calibrate the
sign of \thetaK.

The signs obtained using these three techniques are all consistent
with each other. For example, rotating the PEM counter-clockwise produced a signal with the opposite polarity as applying strain to the ZnSe slide. Furthermore, the signs (determined by rotating the PEM or straining a ZnSe slide) of the transmitted polarization signals produced by a gold film
were consistent with the lower frequency ($<200$~meV) behavior of electrons in a magnetic field.

When the calibrated signs for the measured \thetaF\ and \thetaK\ are used in our thick film
equations, we obtain reasonable optical properties. First, the
real $n$ and imaginary $k$ parts of
the index of refraction are positive. Second, for probe frequencies
below the plasma frequency of metallic samples such as
Au and \sro, the real part of the dielectric function
$\epsilon_1=n^2-k^2$ is negative, implying that $n<k$. The
final signs for \sigmaxx\ and \sigmaxy\ are not arbitrarily
assigned, but are determined from the measurements and the
calibration procedure described here.

\section{Results}

The measurements on the \sro\ and \gma\ films reported here probe
the anomalous Hall effect, which is the Hall effect that arises
from the sample magnetization. Therefore, to eliminate
contributions from the ordinary Hall effect, which depends on the
applied magnetic field, these measurements were performed on films
that are fully magnetized out of plane with zero applied magnetic
field. In the case of \gma\ where the remanent magnetization was very small,
the finite-field linear behavior is extrapolated back to $B=0$.\cite{acbas_icps}
The measurement and analysis techniques described in this
paper can be applied to finite magnetic fields equally well.


Figure~\ref{fig;far_kerr_sro} shows \thetaF\ and \thetaK\ at 10~K
and 0~T from the \sro\ film with the sample fully magnetized out of
plane as a function of probe energy $E$. \thetaF\ and \thetaK\ exhibit strong
energy dependence, with
$\text{Re}\left(\theta\f\right)$ and $\text{Im}
\left(\theta\k\right)$ changing sign at 250~meV and 130~meV, respectively. Below
300~meV, both $\text{Re}\left(\theta\f\right)$ and
$\text{Im}\left(\theta\f\right)$ are negative, indicating that the
change in polarization is in the same sense as the Faraday signals
from free electrons in a gold reference film.  Strictly speaking,
\thetaF\ and \thetaK\ are not defined at zero energy, but they
approach well-defined values as $E\rightarrow 0$. The dc values of
\thetaF\ and \thetaK\ in Fig.~\ref{fig;far_kerr_sro}  are determined
using the thick film equations (Eqs.~(\ref{eq;tanF}) and
(\ref{eq;tanK})) as $E\rightarrow 0$ with the dc measurements for
\sigmaxx\ and \sigmaxy. The dc \thetaH\ at 10~K and 0~T with the
sample fully magnetized is $-0.0052$~rad, which is within 2\% of
$\theta\f(E\rightarrow 0)$ and confirms the expected relationship
for metallic materials where $\theta\f\approx\theta\h$ as frequency goes
to zero.\cite{au} The \thetaF\ and \thetaK\ signals from \sro\
are significantly larger than what has been measured in
non-magnetic metals. For example, in gold, copper, and HTSC films,
where the Faraday signals are produced by free carriers in a
magnetic field, $\text{Re}
\left(\theta\f\right)\approx\text{Im}\left(\theta\f\right)\approx
0.001$~rad at 8~T.\cite{au, cerne_ybco}

The inset of Fig.~\ref{fig;far_kerr_sro} shows \thetaF\ and
\thetaK\ from the \sro\ film as a function of applied magnetic field
at  a probe energy of 117~meV at 10~K.  The response of \thetaF\
to the applied magnetic field is electron-like, as determined by
comparing the signals to a gold film reference sample. Although
the intensity of transmitted light can be as small as 0.01~\%, the magneto-optical signals in transmission
(\thetaF) are approximately an order of magnitude larger than
those obtained in reflection (\thetaK). This is due to the fact
that in metallic films $\theta\f\propto\sigma\xy/\sigma\xx$ while
$\theta\k\propto\sigma\xy/({\sigma\xx})^2$, as is suggested by
Eqs.~\ref{eq;faraday_cond} and \ref{eq;kerr_cond}.  Therefore, for
highly metallic films in the MIR, where $\sigma\xx\gg\sigma\xy$, \thetaF\ is typically larger than \thetaK\ 
for the same \sigmaxy.


Figure~\ref{fig;sigma_sro} shows the measured complex a)
longitudinal conductivity \sigmaxx\ and b) transverse conductivity
\sigmaxy\ for the \sro\ film.  The large symbols are obtained from
\thetaF\ and \thetaK\ measurements at 10~K and and 0~T with the
sample fully magnetized out of plane. The signs of \sigmaxx\ and
\sigmaxy\ represented by the symbols are not assigned arbitrarily,
but are determined experimentally using the techniques described in Section~\ref{sign_sect}. 
The conductivity was defined to include contributions from
both bound and free charges, although the bound charge
contribution in the MIR was found to be small and had almost no
affect on the results.  The heavy solid
($\text{Re}\left(\sigma\xx\right)$) and dashed
($\text{Im}\left(\sigma\xx\right)$) lines in a) are from $H=0$
Kramers-Kronig analysis of reflectance measurements at 35~K and
0~T on a different \sro\ film by Z. Schlesinger's group.\cite{kostic}
The qualitative agreement between the \sigmaxx\
obtained by \thetaF\ and \thetaK\ measurements and that obtained
by reflectance measurements is excellent over the entire energy
range. The quantitative differences could be readily accounted for
by the differences in the two samples, especially since the sample
measured by the Schlesinger group had a dc resistivity at 10~K
of approximately 20~$\mu\Omega$~cm,\cite{kostic, klein3} a
factor of three smaller than the resistivity of the sample used in
the \thetaF\ and \thetaK\ measurements.  The complex \sigmaxx\
obtained using \thetaF\ and \thetaK\ measurements can be compared with values
determined by transmittance and reflectance measurements on the same
sample (small symbols with error bars in
Fig.~\ref{fig;sigma_sro}a).  In the 117-224~meV range,  \sigmaxx\
obtained using \thetaF\ and \thetaK\ measurements is within 20~\%
of the average values obtained using transmittance and reflectance
measurements. At 366~meV the average difference is closer to 30~\%
due to the difficulty in aligning the weak HeNe laser, which has an
output of 2~mW that is two order of magnitude lower than the
CO$_2$ and CO lasers. Furthermore, the differences in the quality
of the polish of the wedged sample and reference substrates are
much more critical at this shorter wavelength. At all wavelengths,
\sigmaxx\ obtained by Faraday and Kerr measurements is within the
error bars of \sigmaxx\ obtained using transmittance and reflectance measurements.
Challenges in absolute transmission and
reflection measurements using discrete laser lines on a wedged
sample could easily account for the differences and suggest that
in this case \thetaF\ and \thetaK\ measurements, which are
self-normalizing, may allow a more accurate determination of
\sigmaxx.


Figure~\ref{fig;far_kerr_gma} shows \thetaF\ and \thetaK\ at 10~K
and 0~T from the \gma\ film with the sample fully magnetized out of
plane as a function of probe energy. \thetaF\ and \thetaK\ show strong
energy dependence, including sign changes. \thetaF\ and \thetaK\
exhibit qualitatively similar behavior in this case.


Figure~\ref{fig;sigma_gma} shows the measured complex a)
longitudinal conductivity \sigmaxx\ and b) transverse conductivity
\sigmaxy\ for the \gma\ film.  These were obtained from \thetaF\ and
\thetaK\ measurements at 10~K and and 0~T with the sample fully
magnetized out of plane. The $E=0$
results are from dc magnetotransport measurements by the Dubon group on similar samples also
grown by them. The symbols with error bars in
Fig.~\ref{fig;sigma_gma}a) shows \sigmaxx\ obtained from
transmittance and reflectance measurements. Since it was more
difficult to obtain an optically smooth polish on GaAs (compared
to \lsgo), and perhaps because it has a larger index of refraction
(compared to \lsgo), differences in the quality of the GaAs
substrate polish led to more than a factor of two uncertainty in
the transmittance at 366~meV due to scattering. Therefore, transmittance and reflectance measurements at 366~meV and shorter wavelengths
are not included. The quality of the
polish made no measurable impact at the longer wavelengths.  Since the Faraday and Kerr measurements involve changes in polarization with applied magnetic field, substrate roughness did not measurably affect the accuracy of the
\thetaF\ and \thetaK\ measurements.
The
transmittance and reflectance measurements led to a \sigmaxx\ that
qualitatively agrees with the \sigmaxx\ obtained by Faraday and
Kerr measurements, which was within the error bars of the
transmittance and reflectance measurements.

The magnitude and frequency dependence of \sigmaxx\ compares
reasonably well with results from other experiments and from
theoretical models. The MIR $\text{Re}\left(\sigma\xx\right)$
extrapolates smoothly to the dc \sigmaxx.
$\text{Re}\left(\sigma\xx\right)$ is similar to that obtained in
Refs.~\onlinecite{burch06} and \onlinecite{singley} for \gma\ samples. The frequency dependence
of $\text{Re}\left(\sigma\xx\right)$ is also similar to that
predicted by theoretical models.\cite{sinova,hankiewicz} One can
also represent the complex conductivity in terms of the complex
dielectric constant $\epsilon$. In this case, the
$\text{Re}\left(\epsilon\right)$ remains at a fairly constant
value of 9 over the measured frequency range whereas
$\text{Im}\left(\epsilon\right)$ strongly decreases with
increasing energy as $E^{-1.3}$, reaching a value of approximately
3 at 0.76~eV. This compares well with ellipsometry measurements on
\gma\ films where $\text{Re}\left(\epsilon\right)$ levels off at a
value between 10 and 12 as $E\rightarrow 0$,
while $\text{Im}\left(\epsilon\right)$ is close 2 at 0.75~eV and
begins to rise as $E\rightarrow 0$.\cite{burch04} Note that the
negative sign in $\text{Im}\left(\sigma\xx\right)$ in
Fig.~\ref{fig;sigma_gma}a) is consistent with the positive
$\text{Re}\left(\epsilon\right)$ found in ellipsometry
measurements in the MIR. The reasonable behavior of \sigmaxx\ in
Fig.~\ref{fig;sigma_gma}a) provides added confidence that the
\sigmaxy\ found in Fig.~\ref{fig;sigma_gma}b) accurately
represents the response of the \gma\ film. The MIR
$\text{Re}\left(\sigma\xy\right)$ extrapolates smoothly to the dc
\sigmaxy. Note that the low energy behavior of the
$\text{Re}\left(\sigma\xy\right)$ is hole-like, suggesting that
there is no sign reversal in the AHE between 0 and 100~meV.

\section{Conclusion}

We have demonstrated experimental and analytical techniques that
can be used to determine the complete complex magneto-conductivity
tensor from Faraday and Kerr measurements in ferromagnetic metals
and semiconductors. It is interesting to note that no absolute intensity measurements
are required; the polarization of the reflected and transmitted light
is sufficient to completely determine the complex magneto-conductivity tensor.
In both \sro\ and \gma\ films, \sigmaxx\ obtained using
Faraday and Kerr measurements was quantitatively consistent with the values for
\sigmaxx\ that were obtained using conventional transmittance and reflectance measurements.
Furthermore, in both materials, \sigmaxy\ showed strong spectral features,
including peaks and sign changes, which will be discussed in future papers.
With the increasing application of
magneto-optical measurements to study magnetic and non-magnetic
materials, and with the specific interest in the infrared
longitudinal and transverse conductivity of magnetic oxides and
semiconductors, these techniques may have a significant impact in
a number of fields.

\begin{acknowledgments}

We wish to thank H.D. Drew for his thick film transmission
calculation, D.C. Schmadel for his advice in constructing our strained ZnSe polarization sign calibrator, and K. Cullinan for his work on designing and machining the translating mount
for the magneto-optical cryostat. We also thank J.S. Dodge for his helpful discussions and John Kielkopf for introducing turcite to us. This work was supported by Research Corporation Cottrell Scholar Award (Buffalo), NSF-CAREER-DMR0449899 (Buffalo), NSF-DMR-0554796 (Santa Cruz), and DOE Contract No. DE-AC03-76SF00098 (Berkeley). Research at Oak Ridge National Laboratory was sponsored by the Division of Materials Sciences and Engineering, Office of Basic Energy Sciences, U.S. Department of Energy, under Contract No. DE-AC05-00OR22725 with Oak Ridge National Laboratory, managed and operated by UT-Battelle, LLC.

\end{acknowledgments}

\vfill\eject

\begin{figure}
\caption{\label{fig;far_kerr_sro} Energy dependence of a) $\theta\f$ and b) $\theta\k$ from a \sro\ film with the sample fully magnetized perpendicular to the plane at 0~T and 10~K. Since the measurement is at $H=0$~T, the OHE as well as background signals from the substrate and windows, which are linear in $H$, do not contribute to the signal. Note the strong energy dependence and the sign changes in $\text{Re}\left(\theta\f\right)$ and $\text{Im} \left(\theta\k\right)$ at 250~K and 130~meV, respectively. The
$\theta\f(E\rightarrow 0)$ plotted in a) is determined using Eq.~\ref{eq;faraday_cond} with the dc measurements for the dc \thetaH\ and \sigmaxx. Inset shows the relationship between
$\text{Re}\left(\theta\f\right)$, $\text{Im}\left(\theta\f\right)$, and $\text{Re}\left(\theta\k\right)$ (which is multiplied by a factor of 5) at 117~meV and 10~K. }
\end{figure}

\begin{figure}
\caption{\label{fig;sigma_sro} The longitudinal conductivity \sigmaxx\ a) and transverse (AHE) conductivity \sigmaxy\ b) for a \sro\ film as a function of probe energy. The solid symbols are obtained from \thetaF\ and \thetaK\ measurements at 10~K and and 0~T with the sample fully magnetized out of plane.
The smaller symbols with error bars in a) are determined from conventional transmittance and reflectance
measurements of the same \sro\ film.
The heavy solid ($\text{Re}\left(\sigma\xx\right)$) and dashed ($\text{Im}\left(\sigma\xx\right)$) lines in a) are from Kramers-Kronig analysis of reflectance measurements at 35~K and 0~T on a different \sro\ sample by Z. Schlesinger's group.}
\end{figure}

\begin{figure}
\caption{\label{fig;far_kerr_gma} Energy dependence of the AHE a) $\theta\f$ and b) $\theta\k$ from a \gma\ film with the sample fully magnetized perpendicular to the plane at 0~T and 10~K. Note the strong energy dependence and the sign changes in both the real and imaginary parts of \thetaF\ and \thetaK. \thetaF\ and \thetaK\ exhibit qualitatively similar behavior in this case.}
\end{figure}

\begin{figure}
\caption{\label{fig;sigma_gma} The longitudinal conductivity \sigmaxx\ a) and transverse (AHE) conductivity \sigmaxy\ b) from a \gma\ film as a function of probe energy. The measurements are at 10~K and and 0~T with the sample fully magnetized out of plane. The smaller symbols with error bars in a) are determined from conventional transmittance and reflectance
measurements of the same \gma\ film.
The $E=0$ results are from dc measurements by the Dubon group on similar samples, which were also grown them.}
\end{figure}

\end{document}